\documentclass[%
 aip,
 amsmath,amssymb,
 reprint,%
 citeautoscript,
]{revtex4-1}

\usepackage{graphicx}%
\usepackage{dcolumn}%
\usepackage{bm}%

\usepackage[utf8]{inputenc}
\usepackage[T1]{fontenc}
\usepackage{mathptmx}
\usepackage{siunitx}
\newcommand{\epsd}{\epsilon^\mathrm{ML}_\mathrm{d}} 
\newcommand{\epso}{\epsilon^\mathrm{ML}_\mathrm{o}}
\newcommand{\qmsbt}{QM7b-T}
\newcommand{\gdbt}{GDB-13-T}
\newcommand{\fchl}{FCHL/$\Delta$-ML}
\newcommand\footnoteref[1]{\protected@xdef\@thefnmark{\ref{#1}}\@footnotemark}
\begin{document}
\title{A Universal Density Matrix Functional from Molecular Orbital-Based Machine Learning: Transferability across Organic Molecules}
\author{Lixue Cheng}
\affiliation{%
Division of Chemistry and Chemical Engineering, California Institute of Technology, Pasadena, CA 91125, USA%
}%
\author{Matthew Welborn}
\affiliation{%
Division of Chemistry and Chemical Engineering, California Institute of Technology, Pasadena, CA 91125, USA%
}%
\author{Anders S. Christensen}
\affiliation{Institute of Physical Chemistry and National Center for Computational Design and Discovery of Novel Materials, Department of Chemistry, University of Basel, Basel, Switzerland}
\author{Thomas F. Miller III}
\email{tfm@caltech.edu.}
\affiliation{%
Division of Chemistry and Chemical Engineering, California Institute of Technology, Pasadena, CA 91125, USA%
}%

\date{\today}%
\begin{abstract}
We address the degree to which machine learning can be used to accurately and transferably predict post-Hartree-Fock correlation energies. Refined strategies for feature design and selection are presented, and the molecular-orbital-based machine learning (MOB-ML) method is applied to several test systems. Strikingly, for the MP2, CCSD, and CCSD(T) levels of theory, it is shown that the thermally accessible (350 K) potential energy surface for a single water molecule can be described to within 1 millihartree using a model that is trained from only a single reference calculation at a randomized geometry. 
To explore the breadth of chemical diversity that can be described, MOB-ML is also applied to a new dataset of thermalized (350 K) geometries of 7211 organic models with up to seven heavy atoms.
In comparison with the previously reported $\Delta$-ML method, MOB-ML is shown to reach chemical accuracy with three-fold fewer training geometries.
Finally, a transferability test in which models trained for seven-heavy-atom systems are used to predict energies for thirteen-heavy-atom systems reveals that MOB-ML reaches chemical accuracy with 36-fold fewer training calculations than $\Delta$-ML (140 versus 5000 training calculations).

\end{abstract}

\maketitle

Machine learning (ML) has recently seen wide application in chemistry, including in the fields of drug discovery,\cite{Lavecchia2015, Gawehn2016, Popova2018} materials design,\cite{Kim2017, Ren2018, Butler2018, Sanchez-Lengeling2018} and reaction prediction.\cite{Wei2016, Raccuglia2016, Ulissi2017,Segler2017, Segler2018} In the context of quantum chemistry, much work has focused on predicting electronic energies or densities based on atom- or geometry-specific features,\cite{Smith2017, smiti_transfer_2018, Lubbers, Bartok2010,rupp2012fast,VonLilienfeld2013,hansen2013assessment, Ceriotti2014,ramakrishnan2015big,Behler2016,kearnes2016molecular,Paesani2016,schutt2017quantum,Tuckerman,Smith2017,wu2018moleculenet,Nguyen2018,Yao2018, Fujikake2018,Li2018,Grisafi2018, Zhang2018}
although other strategies  have also been employed.\cite{mcgibbon2017improving}
Recently, we reported an accurate and transferable molecular-orbital-based machine learning (MOB-ML) approach to the prediction of correlated wavefunction energies based on input features from a self-consistent field calculation, such as the Hartree-Fock (HF) method.\cite{Welborn2018}
In this communication, we present refinements to the MOB-ML method with comparisons to test cases from our previous work.
We then demonstrate the performance of MOB-ML across a broad swath of chemical space, as represented by the QM7b\cite{VonLilienfeld2013} and GDB-13\cite{GDB-13} test sets of organic molecules. 

\section{Theory}
The current work aims to predict post-Hartree-Fock correlated wavefunction energies using features of the Hartree-Fock molecular orbitals (MOs). The starting point for the MOB-ML method\cite{Welborn2018} is that the correlation energy can be decomposed into pairwise occupied MO contributions\cite{Nesbet1958,SzaboNesbet}
\begin{equation}
\label{Ecorr}
E_\textrm{c} 
=\sum^{\textrm{occ}}_{ij}\epsilon_{ij},
\end{equation}
where the pair correlation energy $\epsilon_{ij}$ can be written as a functional of the full set of MOs, $\{\phi_p\}$, appropriately indexed by $i$ and $j$%
\begin{equation}
    \epsilon_{ij} = \epsilon\left[\{\phi_p\}^{ij} \right].
\end{equation}

The functional $\epsilon$ is universal across all chemical systems; for a given level of correlated wavefunction theory, there is a corresponding $\epsilon$ that maps the HF MOs to the pair correlation energy, regardless of the molecular composition or geometry. Furthermore, $\epsilon$ simultaneously describes the pair correlation energy for all pairs of occupied MOs (i.e., the functional form of $\epsilon$ does not depend on $i$ and $j$). 
For example, in second-order M{\o}ller-Plessett perturbation theory (MP2)\cite{Moller1934}, the pair correlation energies are
\begin{equation}
    \epsilon^\textrm{MP2}_{ij} = \frac{1}{4}\displaystyle{\sum_{ab}^\mathrm{virt}}\frac{\left|\left\langle ij \right| \left| ab \right\rangle \right|^2}{e_a + e_b - e_i - e_j}
\end{equation}
where $a$ and $b$ index virtual MOs, $e_p$ is the Hartree-Fock orbital energy corresponding to MO $\phi_p$, and $\left\langle ij \right| \left| ab \right\rangle$ are antisymmetrized electron repulsion integrals.\cite{SzaboNesbet} 
A corresponding expression for the pair correlation energy exists for any post-Hartree-Fock method, but it is typically costly to evaluate in closed form.

In MOB-ML, a machine learning model is constructed for the pair energy functional
\begin{equation}
    \label{eq:ML_functional}
    \epsilon_{ij} \approx \epsilon^\mathrm{ML}\left[\mathbf{f}_{ij}\right]
\end{equation}
where $\mathbf{f}_{ij}$ denotes a vector of features associated with MOs $i$ and $j$.
Eq. \ref{eq:ML_functional} thus presents the opportunity for the machine learning of a universal density matrix functional for correlated wavefunction energies, which can be evaluated at the cost of the MO calculation. %

Following our previous work,\cite{Welborn2018} the features $\mathbf{f}_{ij}$ correspond to unique elements of the Fock ($\mathbf{F}$), Coulomb ($\mathbf{J}$), and exchange ($\mathbf{K}$) matrices between $\phi_i$, $\phi_j$, and the set of virtual orbitals. In the current work, we additionally include features associated with matrix elements between pairs of occupied orbitals for which one member of the pair differs from $\phi_i$ or $\phi_j$ (i.e., non-$i$,$j$ occupied MO pairs).
The feature vector takes the form
\begin{align}
\label{eq:features}
\mathbf{f}_{ij} = & (F_{ii}, F_{ij}, F_{jj},\mathbf{F}^\textrm{o}_{i},\mathbf{F}^\textrm{o}_{j},\mathbf{F}^\textrm{vv}_{ij}, \\ 
 & J_{ii}, J_{ij}, J_{jj},\mathbf{J}^\textrm{o}_{i},\mathbf{J}^\textrm{o}_{j}, \mathbf{J}^\textrm{v}_{i}, \mathbf{J}^\textrm{v}_{j},\mathbf{J}^\textrm{vv}_{ij}, \nonumber\\
 & K_{ij}, \mathbf{K}^\textrm{o}_{i},\mathbf{K}^\textrm{o}_{j},\mathbf{K}^\textrm{v}_{i}, \mathbf{K}^\textrm{v}_{j}
,\mathbf{K}^\textrm{vv}_{ij}). \nonumber
\end{align}
where for a given matrix ($\mathbf{F}$, $\mathbf{J}$, or $\mathbf{K}$) the superscript o denotes a row of its occupied--occupied block, 
the superscript v denotes a row of its occupied--virtual block, 
and the superscript vv denotes its virtual--virtual block.
Redundant elements are removed, such that the virtual--virtual block is represented by its upper triangle and the diagonal elements of $\mathbf{K}$ (which are identical to those of $\mathbf{J}$) are omitted. 
To increase transferability and accuracy, we choose $\phi_i$ and $\phi_j$ to be localized molecular orbitals (LMOs) rather than canonical MOs and employ valence virtual LMOs\cite{Knizia2013IBO} in place of the set of all virtual MOs (as detailed in Ref. \citenum{Welborn2018}).
We separate Eq. \ref{eq:ML_functional} to independently machine learn the cases of $i=j$ and $i\ne j$,
\begin{equation}
    \label{eq:diag_and_offdiag}
    \epsilon_{ij} \approx
    \begin{cases}
        \epsd \left[\mathbf{f}_i\right] & \text{if $i=j$} \\
        \epso \left[\mathbf{f}_{ij}\right] & \text{if $i\ne j$}
    \end{cases}
\end{equation}
where $\mathbf{f}_i$ denotes $\mathbf{f}_{ii}$ (Eq. \ref{eq:features}) with redundant elements removed;
by separating the pair energies in this way, we avoid the situation where a single ML model is required to distinguish between the cases of $i=j$ and $\phi_i$ being nearly degenerate to $\phi_j$, a distinction which can represent a sharp variation in the function to be learned.

In the current work, several technical refinements are introduced to improve training efficiency (i.e., the accuracy and transferability of the model as a function of the number of training examples). These are now described.

\textit{Occupied LMO symmetrization.}
The feature vector is preprocessed to specify a canonical ordering of the occupied and virtual LMO pairs. 
This reduces permutation of elements in the feature vector, resulting in greater ML training efficiency. %
Matrix elements $M_{ij}$ ($\mathbf{M}=\mathbf{F}$, $\mathbf{J}$, $\mathbf{K}$) associated with $\phi_i$ and $\phi_j$ are rotated into gerade and ungerade combinations
\begin{align}
    \label{equation:occupied_sym}
    M_{ii} & \gets \frac{1}{2}M_{ii} + \frac{1}{2}M_{jj} + M_{ij} \\
    M_{jj} & \gets \frac{1}{2}M_{ii} + \frac{1}{2}M_{jj} - M_{ij} \nonumber \\
    M_{ij} & \gets \frac{1}{2}M_{ii} - \frac{1}{2}M_{jj} \nonumber \\
    M_{ip} & \gets \frac{1}{\sqrt{2}}M_{ip} + \frac{1}{\sqrt{2}}M_{jp} \nonumber \\
    M_{jp} & \gets \frac{1}{\sqrt{2}}M_{ip} - \frac{1}{\sqrt{2}}M_{jp} \nonumber
\end{align}
with the sign convention that $F_{ij}$ is negative. Here, $p$ indexes any LMO other than $i$ or $j$ (i.e. an occupied LMO $k$, such that $i \ne k \ne j$, or a valence virtual LMO).

\textit{LMO sorting.} The virtual LMO pairs are sorted by %
increasing distance from occupied orbitals $\phi_i$ and $\phi_j$. Sorting in this way ensures that features corresponding to valence virtual LMOs are listed in decreasing order of heuristic importance, and that the mapping between valence virtual LMOs and their associated features is roughly preserved.
We recognize this issue could also potentially be addressed through the use of symmetry functions,\cite{Behler2007} but these are not employed in the current work.

For purposes of sorting, distance is defined as %
\begin{equation}
R_a^{ij} = \left\Vert \left\langle \phi_i \right| \hat R \left| \phi_i \right\rangle -  \left\langle \phi_a \right| \hat R \left| \phi_a \right\rangle\right\Vert + \left\Vert \left\langle \phi_j \right| \hat R \left| \phi_j \right\rangle - \left\langle \phi_a \right| \hat R \left| \phi_a \right\rangle \right\Vert,
\end{equation}
where $\phi_a$ is a virtual LMO, $\hat R$ is the Cartesian position operator, and $\Vert . \Vert$ denotes the 2-norm.
$\left\Vert \left\langle \phi_i \right| \hat R \left| \phi_i \right\rangle -  \left\langle \phi_a \right| \hat R \left| \phi_a \right\rangle\right\Vert$ represents  the Euclidean distance between the centroids of orbital $i$ and orbital $a$. 
Previously,\cite{Welborn2018} distances were defined based on Coulomb repulsion, which was found to sometimes lead to inconsistent sorting in systems with strongly polarized bonds.
The %
non-$i$,$j$ occupied LMO pairs are sorted in the same manner as the virtual LMO pairs.

\textit{Orbital localization.} 
We employ Boys localization\cite{Boys1960} to obtain the occupied LMOs, rather than IBO localization\cite{Knizia2013IBO} which was employed in our previous work.\cite{Welborn2018}
Particularly for molecules that include triple bonds or multiple lone pairs, it is found that Boys localization provides more consistent localization as a function of small geometry changes than IBO localization;
and the 
chemically unintuitive mixing of $\sigma$ and $\pi$ bonds in Boys localization (``banana bonds'')\cite{Kaldor1967} does not present a problem for the MOB-ML method.

\textit{Feature selection.} Prior to training, automatic feature selection is performed using random forest regression \cite{Breiman2001} with the mean decrease of accuracy criterion (sometimes called permutation importance).\cite{breiman2001statistical} 
This technique was found to be more effective than our previous use\cite{Welborn2018} of the
Gini importance score\cite{Breiman2001}  
which led to worse accuracy and failed to select any features for the case of methane.
 
The reason for using feature selection in this way is twofold. First, GPR performance is known to degrade for high-dimensional datasets (in practice 50-100 features);\cite{Tripathy2016} 
and second, the use of the full feature set with small molecules can lead to overfitting as features can become correlated. 

A reference \textsc{Python} implementation for generating MOB-ML features is provided online. \footnote{See \texttt{https://github.com/thomasfmiller/MOB-ML} for the available code.}

\section{Computational details}
\label{sec:comp}
Results are presented for a single water molecule; a series of alkane molecules; a thermalized version of the QM7b set of 7211 molecules with up to seven C, O, N, S, and Cl heavy atoms; and a thermalized version of the %
GDB-13 set of molecules with thirteen C, O, N, S, and Cl heavy atoms.  All datasets employed in this work are provided in the Supporting Information.

Training and test geometries are sampled at 50 fs intervals from \textit{ab initio} molecular dynamics trajectories performed with the \textsc{Q-Chem} 5.0 software package,\cite{Shao2015} 
using the B3LYP\cite{Vosko1980,Lee1988,Becke1993,Stephens1994}/6-31g*\cite{Hariharan1973} level of theory and a Langevin thermostat\cite{Bussi2007} at 350 K.

The features and training pair energies associated with these geometries are computed using the \textsc{Molpro} 2018.0 software package\cite{MOLPRO} in a cc-pVTZ basis set unless otherwise noted.\cite{Dunning1989} 
Valence virtual orbitals used in feature construction are determined with the Intrinsic Bond Orbital method.\cite{Knizia2013IBO}
Reference pair correlation energies are computed with second-order M{\o}ller-Plessett perturbation theory (MP2)\cite{Moller1934,LMP2} and coupled cluster with singles and doubles (CCSD)\cite{Cizek1966,LCCSD} as well as with perturbative triples (CCSD(T)).\cite{Bartlett1990,LCCSDT}
Density fitting for both Coulomb and exchange integrals~\cite{Polly2004} is employed for all results below except those corresponding to the water molecule.
The frozen core approximation is used in all cases.

Gaussian process regression (GPR)\cite{rasmussen2006} is employed to machine learn $\epsd$ and $\epso$ (Eq. \ref{eq:diag_and_offdiag}) using the \textsc{GPy} 1.9.6 software package. \cite{gpy2014}
The GPR kernel is Mat\'ern 5/2 with white noise regularization\cite{rasmussen2006}. Kernel hyperparameters are optimized with respect to the log marginal likelihood objective for the water and alkane series results, as well as for $\epsd$ of the QM7b results. We use the Mat\'ern 3/2 kernel instead of the Mat\'ern 5/2 kernel for the case of $\epso$ for QM7b results, as it was empirically found to yield slightly better accuracy.
\footnote{\label{github}In principle, the smoothness of the Mat\'ern kernel could be taken as a kernel hyperparameter; however, this possibility was not explored in this work.} %

Feature selection is performed using the random forest regression implementation in the \textsc{Scikit-learn} v0.20.0 package. \cite{scikit-learn} 

\section{Results}
The ML model of Eq.~\ref{eq:diag_and_offdiag} is a universal functional for any molecular Hamiltonian.  In principle, with an adequate feature list and unlimited training data (and time), it should accurately and simultaneously describe all molecular systems.
In practice, we must train the ML model using a truncated feature list and finite data. These choices determine the accuracy of the model. 

Below, we examine the performance of the MOB-ML method in three increasingly broad regions of chemical space:
(i) training on randomized water molecule geometries and predicting the energies of other water molecule geometries;  
(ii) training on geometries of short alkanes and predicting the energies of longer alkanes, 
and (iii) training on a small set of organic molecules and predicting the energies of a broader set of organic molecules.
The first two test cases were introduced in our previous work,\cite{Welborn2018} and  we explore how the refined methodology reported herein leads to improvements in accuracy and transferability.
The last case represents a demanding new test of transferability across chemical space.
In all cases, we report the ML prediction accuracy as a function of the number of training examples. 

\begin{figure}[htbp]
\includegraphics[width=1.0\columnwidth]{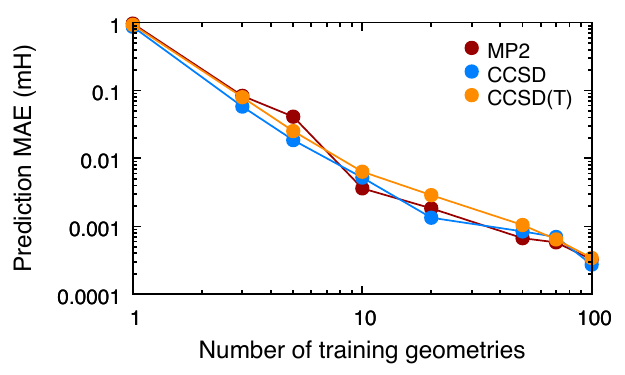}
\caption{
Learning curves for MOB-ML models trained on the water molecule and used to predict the correlation energy of different water molecule geometries at three levels of post-Hartree-Fock theory. 
Prediction errors are summarized in terms of mean absolute error (MAE).
}
\label{figure:water}
\end{figure}

As a first example, we consider the performance of MOB-ML for a single water molecule. 
A separate model is trained to predict the correlation energy at the MP2, CCSD, and CCSD(T) levels of theory, using reference calculations on a subset of 1000 randomized water geometries to predict the correlation energy for the remainder.
Feature selection with an importance threshold of \num{1e-3} results in 12, 11 and 10 features for $\epso$ for MP2, CCSD and CCSD(T), respectively; 
ten features are selected for $\epsd$ for all three post-Hartree-Fock methods. 
Figure \ref{figure:water} presents the test set prediction accuracy of each MOB-ML model as a function of the number of training geometries (i.e., the ``learning curve"). 
MOB-ML predictions are shown for MP2, CCSD, and CCSD(T), and the model shows the same level of accuracy for all three methods. 
Remarkably, all three models achieve a prediction mean absolute error (MAE) of 1 mH when trained on only a single water geometry, indicating that only a single reference calculation is needed to provide chemical accuracy for the remaining 999 geometries at each level of theory. 
Since it contains 10 distinct LMO pairs, this single geometry provides enough information %
to yield a chemically accurate MOB-ML model for the global thermally accessible potential energy surface.

For all three methods (Fig.~\ref{figure:water}), the learning curve exhibits the expected\cite{learning_curves} power-law behavior as a function of training data, and the total error reaches microhartree accuracy with tens of water training geometries.
As compared to our previous results, where training on 200 geometries resulted in a prediction MAE of 0.027 mH for the case of CCSD,\cite{Welborn2018}
the current implementation of the MOB-ML model is substantially improved; the improvement for this case stems primarily from the use of Boys localization,\cite{Boys1960} which specifies unique and consistent LMOs corresponding to the oxygen lone pairs.

Next, we explore the transferability of MOB-ML predictions for a model that is trained on thermalized geometries of short  alkanes and then used  for predictions on thermalized geometries of larger and more branched alkanes (n-butane 
and isobutane).
For these predictions, the absolute zero of energy is shifted for each molecule to compare relative energies on its potential energy surface (i.e., parallelity errors are removed). These shifts are reported in the figure caption; for no other results reported in the paper are parallelity errors removed.

\begin{figure}[htbp]
\includegraphics[width=1.0\columnwidth]{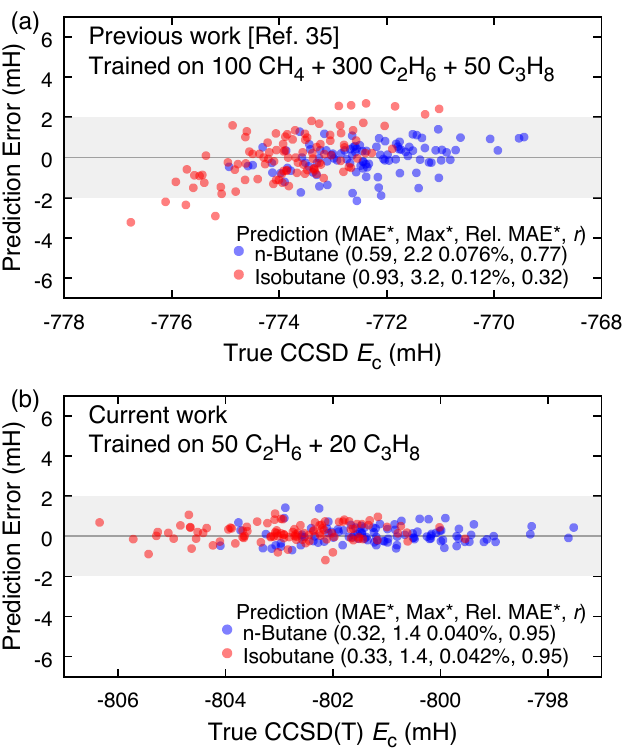}
\caption{
MOB-ML predictions of the correlation energy for 100 n-butane and isobutane geometries, using MOB-ML features described in the (b) current work , compared to (a) the previous MOB-ML features of Ref. \citenum{Welborn2018}.
Training sets are indicated in each panel of the figure.
MOB-ML prediction errors are plotted versus the (a) true CCSD and (b) true CCSD(T) correlation energy. To remove parallelity error, a global shift is applied to the predictions of n-butane and isobutane by (a) 3.3 and 0.73 mH and (b) 0.90 and 0.17 mH, respectively. Summary statistics that include this shift (indicated by an asterisk) are presented, consisting of mean absolute error (MAE*), maximum absolute error (Max*), MAE* as a percentage of $E_\mathrm{c}$ (Rel. MAE*), and Pearson correlation coefficient ($r$)\cite{Pearson1896}. The gray shaded region corresponds to errors of $\pm 2$ mH.
}
\label{figure:alkanes}
\end{figure}

In our previous work,\cite{Welborn2018} this test was performed using training data that combined of 100 geometries of methane, 300 of ethane, and 50 or propane; the resulting predictions are reproduced here in Fig.~\ref{figure:alkanes}a.  This earlier implementation of MOB-ML led to predictions for n-butane 
and isobutane with substantial errors (0.59 mH for n-butane and 0.93 mH for isobutane) and noticable skew with respect to the true correlation energy.

The predictions of MOB-ML in the current work (Fig.~\ref{figure:alkanes}b) are markedly improved.
First, the overall prediction accuracy is improved for all four summary statistics (inset in Fig. \ref{figure:alkanes}) despite substantial reduction in the number of training examples used. (The current work uses only 50 geometries of ethane, 20 geometries of propane, and no methane data.) Second, n-butane and isobutane are predicted with nearly identical accuracy. Finally, the prediction errors are no longer skewed as a function of true correlation energy. 
The primary methodological sources of these improvements are found to be symmetrization of occupied orbitals (Eq.~\ref{equation:occupied_sym}) and the improved feature selection methodology. The MOB-ML features in the current work are selected with an importance threshold of \num{1e-4}, resulting in  27 features for $\epsd$ and 12 features for $\epso$; results presented in Fig \ref{figure:alkanes}b for CCSD(T) are qualitatively identical to those obtained for CCSD (not shown).

We now examine the transferability of the MOB-ML method across a broad swath of chemical space.
Specifically, we consider the QM7b dataset,\cite{QM7b} which is comprised of 7,211 plausible organic molecules with 7 or fewer heavy atoms. 
The chemical elements in QM7b are limited to those likely to be found in drug-like compounds: C, H, O, N, S, and Cl. 
We refer to the dataset used herein as \qmsbt{} to reflect the fact that it contains geometries sampled at a temperature of 350 K (as described in Sec. \ref{sec:comp}), as opposed to DFT-optimized geometries. 
The MOB-ML model is trained on a randomly chosen subset of \qmsbt{} molecules and used to predict the correlation energy of the remainder. 
Active learning was also tested as a training data selection strategy, but was not found to improve the predictions in the regime of chemical accuracy, and in fact led to slightly worse transferability.

For comparison, a $\Delta$-ML model\cite{ramakrishnan2015big} was trained on the same molecules using kernel-ridge regression using the FCHL representation\cite{Faber2018} with a Gaussian kernel function (\fchl), as implemented in the QML package.\cite{Christensen2019}
All hyperparameters of the model were set to those obtained in Ref. \citenum{Faber2018}, which have previously been demonstrated to work well for datasets containing structures similar to those in \qmsbt{}.\cite{Christensen2019} 

A possible source of concern for MOB-ML is that the number of selected features would grow with the chemical complexity of the training data. 
For example, 27 features for $\epsd$ and 12 features for $\epso$ were selected in the alkane test case using ethane + propane training data (Fig. \ref{figure:alkanes}b), whereas only 10 features for $\epsd$ and 10 features for $\epso$ were selected for the water test case at the CCSD(T) level of theory (Fig. \ref{figure:water}).
To examine this, we perform feature selection on increasing numbers of randomly selected molecules from the \qmsbt{} dataset. 
Table \ref{table:qm7b_features} presents two statistics on the feature importance as a function of the number of training molecules: (i) the number of ``important features" (i.e., those whose permutation importance\cite{breiman2001statistical} exceeds a set threshold of \num{2e-4} and \num{5e-5} for $\epsd$ and $\epso$, respectively) and (ii) the inverse participation ratio\cite{IPR} of the feature importance scores. The latter is a threshold-less measure of the number of important features; it takes a value of 1 when only 1 feature has nonzero importance and $N$ when all $N$ features have equal importance. 
Although the \qmsbt{} dataset contains many different chemical elements and bonding motifs, Table \ref{table:qm7b_features} reveals that the selected features remain compact and do not grow with the number of training molecules. Indeed, for a large number of training molecules, the number of selected features slightly decreases, reaching 42 and 24 selected features for $\epsd$ and $\epso$, respectively, for the largest training sizes considered. 

\begin{table}[htbp]
\caption{Number of features selected as a function of the number randomly chosen  training molecules for the \qmsbt{} dataset at the CCSD(T)/cc-pVDZ level. 
The number of features that exceed an importance threshold as well as the inverse participation ratio (IPR) of the feature importance scores are reported (see text).
}
\label{table:qm7b_features}
\begin{ruledtabular}
\begin{tabular}{lrrcccc}
&\multicolumn{2}{c}{\# of important features}  &\multicolumn{2}{c}{feature weight IPR}\\
Training size&$\epsd$&$\epso$&$\epsd$&$\epso$ \\ \hline
20 & 50 & 28 & 4.720 & 1.116 \\
50 & 46 & 28 & 3.718 & 1.097\\
100 & 46 & 26 & 3.450 & 1.115\\
200 & 42 & 24 & 3.430 & 1.120\\
\end{tabular}
\end{ruledtabular}
\end{table}

\begin{figure}[htbp]
\includegraphics[width=0.95\columnwidth]{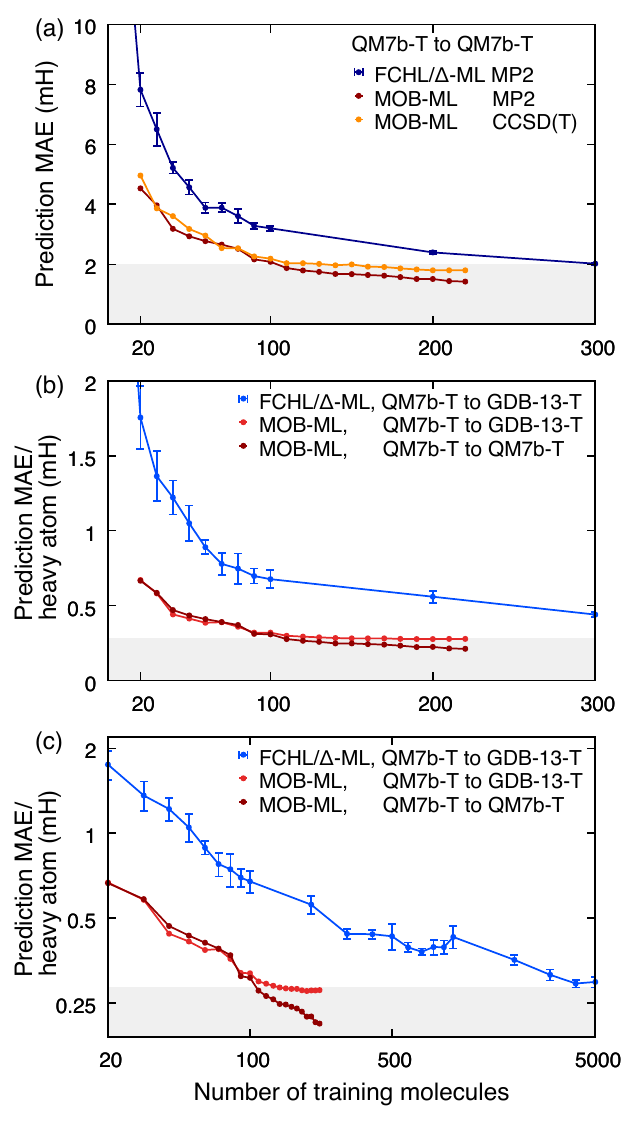}
\caption{
Learning curves for MOB-ML trained on \qmsbt{} and applied to \qmsbt{} and \gdbt{} (see text for definition of these datasets).
\fchl \cite{Faber2018} results are provided for comparison. 
(a) Predictions are made for \qmsbt{} at the MP2/cc-pVTZ (red) and CCSD(T)/cc-pVDZ (orange) levels of theory. 
(b) Using the same models trained on \qmsbt, predictions are made for \gdbt, and reported in terms of MAE per heavy atom. (MOB-ML predictions for \qmsbt{} are included for reference.)
(c) As in the previous panel, but plotted on a logarithmic scale and extended to show the full range of \fchl{} predictions.
Error bars for \fchl{} represent prediction standard errors of the mean as measured over 10 models.
The gray shaded area corresponds to errors of 2 mH per 7 heavy atoms.
}
\label{figure:qm7b_comparison}
\end{figure}

The  learning curves for MOB-ML models trained at the MP2/cc-pVTZ and CCSD(T)/cc-pVDZ levels of theory
are shown in Fig.~\ref{figure:qm7b_comparison}a, as well as the \fchl{} learning curve for MP2/cc-pVTZ. 
At the MP2 level of theory, the MOB-ML model achieves an accuracy of 2 mH with 110 training calculations (representing 1.5\% of the molecules in the \qmsbt{} dataset), whereas the \fchl{} requires over 300 training geometries to reach the same accuracy threshold.
Fig.~\ref{figure:qm7b_comparison}a also illustrates the relative insensitivity of MOB-ML to the level of electronic structure theory, with the learning curve for CCSD(T)/cc-pVDZ reaching 2 mH accuracy with 140 training calculations. 
An analysis of the sensitivity of the MOB-ML predictions to the number of selected features is presented in supporting information Fig. S1, which indicates that the reported results are robust with respect to the number of selected features.

As a final test of transferability of the MOB-ML and \fchl{} methods across chemical space, Figs. \ref{figure:qm7b_comparison}b and \ref{figure:qm7b_comparison}c show results in which the ML methods are trained on \qmsbt{} molecules and then used to predict results for a  dataset of 13-heavy-atom organic molecules at thermalized geometries, \gdbt, which includes six thermally sampled geometries each of 1,000 13-heavy-atom organic molecules chosen randomly from the GDB-13 dataset.\cite{GDB-13}
Like QM7b, the members of GDB-13 contain C, H, N, O, S, and Cl. 
The size of these molecules precludes the use of coupled cluster theory to generate reference data; we therefore make comparison at the MP2/cc-pVTZ level of theory, noting that MOB-ML has consistently been shown to be insensitive to the employed post-Hartree-Fock method (as in Fig. \ref{figure:qm7b_comparison}a). 
Transfer learning results as a function of the number of training molecules are presented in Figs. \ref{figure:qm7b_comparison}b (on a linear-linear scale) and \ref{figure:qm7b_comparison}c (on a log-log scale).

Using the MOB-ML model that is trained on 110 seven-heavy-atom molecules (corresponding to a prediction MAE of 1.89 mH for \qmsbt), we observe a prediction MAE of 3.88 mH for \gdbt.
Expressed in terms of size-intensive quantities, the prediction MAE per heavy atom is 0.277 mH and 0.298 mH for \qmsbt{} and \gdbt, respectively, indicating that the accuracy of the MOB-ML results are only slightly worse when the model is transferred to the dataset of larger molecules. 
On a per-heavy-atom basis, MOB-ML reaches chemical accuracy with the same number of  \qmsbt{} training calculations (approximately 100), regardless of whether it is tested on \qmsbt{} or \gdbt.

In contrast with MOB-ML, the \fchl{} method is found to be significantly less transferable from  \qmsbt{} to \gdbt.  For models trained using 100 seven-heavy-atom molecules, the MAE per heavy atom of \fchl{} is over twice that of MOB-ML (Fig. \ref{figure:qm7b_comparison}b).  Moreover, whereas MOB-ML reaches the per-heavy-atom chemical accuracy threshold with 140 training calculations, the \fchl{} method only reaches that threshold with 5000 training calculations.

\section{Conclusions}

Molecular-orbital-based machine learning (MOB-ML) has been shown to be a simple and strikingly accurate strategy for predicting correlated wavefunction energies at the cost of a Hartree-Fock calculation, benefiting from the intrinsic transferability of the localized molecular orbital representation.
The starting point for the MOB-ML method is a rigorous mapping from the Hartree-Fock molecular orbitals to the total correlation energy, which ensures that the use of sufficient training data and molecular orbital features will produce a model that matches the corresponding correlated wavefunction method across the entirety of chemical space.
The current work explores this possibility within the subspace of organic molecules. %
It is shown that MOB-ML predicts energies of the \qmsbt{} dataset to within a 2 millihartree accuracy using only 110 training calculations at the MP2/cc-pVTZ level of theory and using 140 training calculations at the CCSD(T)/cc-pVDZ level of theory. 
Direct comparison with \fchl{} reveals that MOB-ML is threefold more efficient in reaching chemical accuracy for describing \qmsbt. Further, a transferability test of a MOB-ML model trained on \qmsbt{} to \gdbt{} reveals that MOB-ML exhibits negligible degradation in accuracy; as a result, chemical accuracy is achieved with 36-times fewer training calculations using MOB-ML versus \fchl.
These results suggest that MOB-ML %
provides a promising approach toward the development of density matrix functionals that are applicable across broad swathes of chemical space.

\begin{acknowledgments}
We thank Daniel Smith (Molecular Sciences Software Institute) and Alberto Gobbi (Genentech) for a helpful discussion about available training datasets.
T.F.M. acknowledges support from AFOSR award no. FA9550-17-1-0102. A.S.C. acknowledges support from 
the National Centre of Competence in Research (NCCR) Materials Revolution: Computational Design and Discovery of Novel Materials (MARVEL) of the Swiss National Science Foundation (SNSF).
We additionally acknowledge support from the Resnick Sustainability Institute postdoctoral fellowship (M.W.) and the Camille Dreyfus Teacher-Scholar Award (T.F.M.).
Computational resources were provided by the National Energy Research Scientific
Computing Center (NERSC), a DOE Office of Science User Facility
supported by the DOE Office of Science under contract DE-AC02-05CH11231.
\end{acknowledgments}

\section*{Supporting Information}
The datasets used in this work are available for download;\cite{dataset} they include MOB-ML features, HF energies, pair correlation energies, and geometries.
MOB-ML and \fchl{} predictions corresponding to Fig. \ref{figure:qm7b_comparison} and an analysis of the sensitivity of the results of Fig. \ref{figure:qm7b_comparison} to the number of selected features are available in the attached supporting information. 
A reference \textsc{Python} implementation for generating MOB-ML features is available at \texttt{https://github.com/thomasfmiller/MOB-ML}.%

\bibliography{main}%
\end{document}